\documentclass[aps,prl,twocolumn]{revtex4}

\usepackage{amsmath,amsthm}
\usepackage{graphicx}
\usepackage{flafter}
\usepackage{subfig}
\usepackage{setspace}

\begin{document}

\title{Meson coupling constants at high mass and large $N_c$}

\author{Thomas D. Cohen}
\email{cohen@physics.umd.edu}

\author{Elizabeth S. Werbos}
\email{ewerbos@physics.umd.edu}

\affiliation{Department of Physics, University of Maryland,
College Park, MD 20742-4111}

\begin{abstract}

Quark-hadron duality implies that a process described in terms of
quark loops should be the hadronic amplitude when averaged over a
sufficient number of states.  Ambiguities associated with the
notion of quark hadron duality can be made arbitrarily small for
highly excited mesons at large $N_c$.   QCD is expected to form a
string like description at large $N_c$ yielding an exponentially
increasing Hagedorn spectrum for high mass.  It is shown that in
order to reconcile quantum-hadron duality with a Hagedorn
spectrum, the magnitude of individual coupling constants between
high-lying mesons in a typical decay process must be
characteristically larger than the average of the coupling
constants to mesons with nearby masses.  The ratio of the square
of the  average coupling to the average of the coupling squared
(where the average is over mesons with nearby masses) drops {\it
exponentially} with the mass of the meson. Scenarios are discussed
by which such a high precision cancellation can occur.

\end{abstract}

\maketitle

It is well known that QCD in the limit of an infinite number of
colors has an infinite number of infinitely narrow meson
states\cite{Witten}. The narrowness arises from the fact that
meson-meson interactions are suppressed by powers of $1/N_c$.
Clearly at finite $N_c$ the widths are finite due to decays. At
large but finite $N_c$, one expects that mesons of increasing mass
become increasingly broad---eventually to the point where they are
no longer discernable.  It is not immediately clear what, if
anything, one can deduce about multi-meson couplings from general
considerations of large $N_c$ QCD.  In this paper we show that we
can infer important qualitative information about couplings of
highly excited mesons at large but finite $N_c$.  In particular, one
can show that the magnitude of the coupling constant for a typical
three meson vertex involved in the decay process is much larger
than the average of such coupling over a large number of mesons
with the same quantum numbers and nearby masses.  Indeed, at large
$N_c$ the ratio of the square of the average coupling to the
average of the coupling squared (where the average is over mesons
with nearby masses) drops {\it exponentially} with the mass of the
meson.

This rather striking result can be derived from two well-founded
pieces of physics which are thought to become exact at large $N_c$.
The first of these is quark-hadron duality---the general principle
that the amplitude for a hadronic process smeared over a sufficient
number of states will be equal to the amplitude as calculated from a
(perturbative) quark process\cite{Bloom}.  The second is the widely
accepted view that at large $N_c$ and high excitations mesons can be
represented by QCD strings\cite{Pol,Rebbi}.  We should note here that
analogous arguments can be formulated for glueballs and one expects
the same qualitative features.

We wish to stress, that result here is by no means obvious.  Clearly,
a Hagedorn spectrum implies that any two-point correlation function of
QCD composite operators will have couplings of the external couplings
falling exponentially with the mass: the spectral strength is simply
shared by many states.  For similarly pedestrian reasons, the
three-point coupling averaged over many nearby hadrons can also be
shown to fall exponentially rapidly with mass. What is remarkable
here, is not that average value of the three-point coupling is small,
but rather that the small average value does {\it not} imply that the
typical three-point coupling for any given decay is also small.  In
fact the typical coupling is radically larger than its average.  Thus,
there must be extraordinary correlations between the individual
couplings to yield such a small average coupling.

Before proceeding with a detailed description of the present
problem, a few general comments about the nature of the large
$N_c$ limit, quark hadron duality, and the QCD string are in
order.  One central issue is that of ordering; as is common in
large $N_c$ QCD\cite{TomsLimits}, ordering of limits can play a
critical role.  In the present case, the key point is that if one
first considers the limit of high mass followed by the large $N_c$
limit, the spectrum for some correlation function is described by
a QCD continuum without discernable resonances.  In contrast, if
one first takes the large $N_c$ limit with the large mass limit,
then the spectrum will contain infinitely narrow resonances.  The
difference between these implies that the double limit will not
converge uniformly for the spectrum. Here we will be focusing on a
regime in which $N_c$ is large enough so that the mesons under
consideration are weakly interacting and narrow enough to be
treated as well isolated resonances.  In practice this means that
widths have to be narrow enough compared to the spacing between
mesons of fixed quantum number. At the same time the masses under
consideration must be high enough so that a)the mass scale is
perturbative in the sense that it is large compared with
$\Lambda_{\rm QCD}$, and b) the density of states for mesons
contains states well into the Hagedorn region of exponential
growth. The tension between the two limits is implicit here.  It
is easy to see that as a problem in mathematical physics one ought
to be able to satisfy these conditions with arbitrary accuracy;
the extent to which such a regime will turn out to be of relevance
for the $N_c=3$ world is a bit more problematic.

The notion of quark hadron duality can be a bit elusive; formulating
it in a precise mathematical way may be quite difficult when modelling
a specific process. While the underlying idea is clear---that the two
descriptions should become equivalent when averaged over some number
of states---it is often unclear how many states need to be included
\cite{Isgur}. However, these ambiguities should disappear when going
to highly excited states at large $N_c$. To make things concrete we
will focus on the correlation function of local currents which are
expressible in terms of quark (and/or gluon) operators.  General
arguments based on asymptotic freedom and the operator product
expansion imply that the perturbative quark loop will dominate
correlation functions in the exact QCD expression provided that one
studies them at large virtuality\cite{pgw}. It is generally believed
that this is consistent only if ``semi-local'' duality holds in the
sense that averages over the full spectral function are over mass
scales large compared with $\Lambda_{QCD}$\cite{szv,Narison}.  This is
potentially problematic for the physical world in that this might
force one to average over a region where discrete mesons have already
``melted'' into the QCD continuum. By considering a world with
sufficiently large $N_c$ we can ensure that this problem does not
occur.  Moreover, at large $N_c$ there is no difficulty in considering
sufficiently high-lying states so that the mass scale is unambiguously
perturbative while at the same time having clearly discernable
hadronic states\cite{cohen&glozman,Shifman}.

There are two critical aspects of the QCD string which play a role
in the present analysis.  The first is the existence of  the
Hagedorn spectrum for the density of meson states at high
excitation\cite{Hag,Pol,Dien}:
\begin{equation}
\label{HagSpectrum} \rho(m) = A(M/T_H)^{-2B} e^{M/T_H} \textrm{
(A,B constant)} \; .
\end{equation}
The second is that the decay width for a meson as represented by a
string should be proportional to the mass of the meson
(corresponding to the length of the string). This is essentially a
uniform probability for the string to break per unit
length\cite{Nuss}.  Combining this with the
fact that the decay amplitude for a meson breaking into two mesons
goes as $N_c^{-1/2}$ implies that
\begin{equation}
\label{lnsdw}
\Gamma_s(m) \sim \frac{1}{N_c}\frac{\Lambda}{\sigma} m \; ,
\end{equation}
where $\Lambda$ is the QCD scale and $\sigma$,  the string
tension, is independent of $N_c$.

The strategy employed here is straightforward.  We consider a
three-point correlation function for currents with mesonic quantum
numbers.  Working in the regime discussed above, we evaluate the
correlation function two ways: i) as a quark loop, and ii) as the
sum over narrow mesons.  We extract the spectral strength of the
correlation function averaged over a range of masses for the two
descriptions.  By standard OPE type arguments the first
description should be valid at sufficiently large masses and a
sufficiently large averaging region.  The meson-based description
should be valid at sufficiently large $N_c$ and depends on unknown
coupling constants. Since there is a common region of validity of
the two descriptions, they should match.  To proceed further we
parameterize the coupling constants in the meson description by an
average value over a large number of states times a factor which
is state specific.  The average value is determined by the
matching on to the quark description.

Next we consider the width of a decaying meson based on the
assumption that a typical coupling constant is approximately given
by the average.  The width so computed depends only on known
quantities and the density of states.  If one takes the density of
states to be Hagedorn-like as one expects in a string theory, then
one finds that width decreases exponentially with the mass of the
meson.  However, this violates the expectation that in a string
description of hadrons the decay width grows linearly with the
mass as given in Eq.~\ref{lnsdw}.  Thus we conclude that the
assumption that a typical coupling contributing to the decay is
equal to the average coupling is incorrect.

To achieve consistency between the string description and quark
hadron duality at large $N_c$ there must be very large
cancellations of some sort: the typical coupling relevant in the
decay will be exponentially larger than the average.  To make this
more precise, the ratio of the square of the average coupling to
the average of the coupling squared drops {\it exponentially}
with the mass of the meson.  In defining this the phases of the
coupling are fixed by requiring that the meson field has the same
phase as the current acting on the vacuum.

There are some illuminating observations we can make about the
structure of a three-point meson interaction. Our concern is the
impact that the Hagedorn spectrum has on its form.

We consider a process with the physical interpretation that the
current $A$ creates a meson of type $a$ which then decays into mesons
of type $b$ and $c$ which are destroyed by the analogous $B$ and $C$
currents (see eq. (\ref{eqn:threeptfirst}) in the Appendix for a
precise mathematical formulation of the amplitude).

The first thing we observe is that the amplitudes for $A$, $B$, and
$C$ to create mesons will also be part of the two-point function, and
as such their averages can be determined by comparing the two-point
function to the analogous quark loop.

For the comparison with the quark loop, we can exploit dimensional
analysis to simplify the issues. When we calculate both
descriptions of the scattering amplitudes over a small region of
momentum and invoke duality to calculate the decay width for the
meson, the scale dependence has two sources: the quark loop and
the density of meson states. When the mass is very large compared
to $\Lambda_{QCD}$ the only scales playing a role in the spectral
function for the quark loop arise from the four-momenta of the
mesons. When one integrates over final states to find the total
decay width for a meson state, then the only scales left are $m$,
the mass of the initial meson in its rest frame, and any scales
which come into play from the density of mesonic states.

Keeping all this in mind, we find for the two point function from the
quark loop with two insertions:
\begin{equation}
 \overline{a_{\lambda}}(s) = \sqrt{\frac{Q_{2 a} (s,\lambda)}{\rho_{a \lambda}(s)}}
\end{equation}
where $\overline{a_{\lambda}}(s)$ is the average value for a particle
of a state $\lambda$ to be created with a momentum $s$, $Q_{2 a}
(s,\lambda)$ is the amplitude of the quark loop with two insertions,
and $\rho_{a \lambda}(s)$ is the density of states (See eq. (\ref{eqn:rhodef})-(\ref{eqn:abardef}) in the Appendix for precise definitions).

Thus, the amplitude to create any particular state must decrease as
the density increases, as the quark loop is not a function of this
density, but only of the kinematic parameters and the strong coupling.

We can do the same thing in the case of the three-point function, and
find:
\begin{equation}
\begin{split}
Q_3'\left(s_0,\lambda_0,\dots\right) & =
\\ & \overline{f}\left(s_0,\lambda_0,..\right)
\left(\rho_{a \lambda_0}(s_0) \overline{a_{\lambda_0}}(s_0)\right)
\\ & \times      \left(\rho_{b \lambda_1}(s_1) \overline{b_{\lambda_1}}(s_1)\right)
   \left(\rho_{c \lambda_2}(s_2) \overline{c_{\lambda_2}}(s_2)\right)
\end{split}
\end{equation}
where $Q_3'$ is the spectral strength of the quark three point
function as defined in eq. (\ref{eqn:Q3def}) in the Appendix. $f$ is
an unknown function proportional to the three-point coupling which we
can use to compute a meson decay width, and $\overline{f}$ is its
average (as defined in eq. (\ref{eqn:fbardef}) in the Appendix). This
equation, along with what we know from the two-point function, gives
us its form as:
\begin{equation}
\begin{split}
\overline{f} = & \frac{1}{\sqrt{\rho_{a \lambda_0}(s_0)
                                \rho_{b \lambda_1}(s_1) \rho_{c
                                \lambda_2}(s_2)}} \\ & \times
                                \frac{Q_3'\left(s_0,s_1,s_2\right)}{\sqrt{Q_{2a}\left(s_0\right)
                                Q_{2b}\left(s_1\right)
                                Q_{2c}\left(s_2\right)}}
\end{split}
\end{equation}

The next step is to relate $f$ to the decay width for a specific
meson.

We have determined $\overline{f}$ from a relationship with the quark
loop. However, the $f$ in each meson's decay width is specific to the
individual meson, and is not guaranteed to be the same as its
average. Anticipating that it will not be, we quantify this difference
by defining:
\begin{equation}
f \equiv R \overline{f}
\end{equation}
R quantifies the unknown behavior of the individual meson coupling
constants, and is thus far wholly unconstrained.

To find the total cross section for the decay of a particle of mass
$m_0$, we integrate over all mass states and sum over all spin states
for the outgoing particles. The density of states in this integration
exactly cancels the density dependence from the coupling constant for
the final states, so the full decay width is then:
\begin{equation}
\begin{split}
\Gamma(m_0,\lambda_0) = \frac{1}{\rho_{a \lambda_0}(m_0)}
    \sum_{\lambda_1,\lambda_2} \iint & dm_1 dm_2
             |R|^2
\\ & \times      Q(m_0,\lambda_0,..)
\end{split}
\end{equation}
Here, $Q$ includes factors from the quark loop and kinematic
information. Its exact form (eq. (\ref{eqn:Qdef}) in the Appendix) is
unimportant--The key point is its mass dependance.

We now make the observation that, aside from $\rho_{a \lambda_0}(m_0)$
and $R$, the only mass scale in the problem is $m_0$. As for $N_c$,
the quark loops yield the correct dependence of
$\frac{1}{N_c}$\cite{Witten}.

If it were the case that $R=1$, meaning the average coupling constant
was equal to each individual coupling constant, the decay width would
be:
\begin{eqnarray}
\Gamma(m_0,\lambda_0) & = & \frac{1}{N_c}\frac{1}{\rho_{a \lambda_0}(m_0)} F(m_0)
\\F(m_0)& = & \sum_{\lambda_1,\lambda_2} \iint  dm_1 dm_2
              Q(m_0,\lambda_0,..)
\end{eqnarray}
The density function has units of $m^{-1}$, so $\frac{1}{\rho_{a
\lambda_0}(m_0)}$ has units of $m$--the same units as the decay
width. Therefore, in this case, $F(m_0)$ would be dimensionless,
and, since it depends only on one scale, completely independent of
this scale (but only at an energy much larger than $\Lambda_{QCD}$
and the quark masses).

If, as is generally believed, the density of states follows the
Hagedorn spectrum in Eq. (\ref{HagSpectrum}), this width must
exponentially decrease with mass. In contradiction, string theory
predicts a decay width linearly increasing with mass as given in
Eq.~(\ref{lnsdw}).

It is clear, then, that $R$ must have a highly nontrivial mass
dependence. We cannot derive the full form of $R$ using these
general methods, as it can depend on $m_0$, $m_1$ and $m_2$. For
illustration of the effect we will set $R = R(m_0)$, ignoring any
possible dependence on the other two masses. In this case,
\begin{equation}
\Gamma(m_0,\lambda_0) = \frac{1}{N_c}\frac{|R(m_0)|^2}
                                          {\rho_{a \lambda_0}(m_0)}
                        F(m_0)
\end{equation}
$F$ is still independent of $m_0$ by the above argument, which means:
\begin{equation}
\frac{\Gamma}{\Gamma_s} \sim \frac{|R(m_0)|^2}{m_0 \rho_{a \lambda_0}(m_0)}
                         F \frac{\sigma}{\Lambda}
\end{equation}
Therefore, if $R$ were only a function of $m_0$, its form would be:
\begin{equation}
|R(m_0)|^2 \sim m_0 \rho_{a \lambda_0}(m_0)
\end{equation}
Using the Hagedorn spectrum in Eq. (\ref{HagSpectrum}), $\rho_{a
\lambda_0}(m_0)$ is exponentially increasing and $|R(m_0)|^2$ must
also be exponentially increasing.

This simple calculation demonstrates that individual meson
coupling constants must differ substantially from their average
over some small momentum-squared region. The integration range is
large enough to contain sufficient states for duality to be valid,
but small enough that the amplitude for the quark loop (which we
know to be smooth) can be taken to be constant.

In essence, then, the coupling constants within any
momentum-squared range large enough for duality to be valid must
have exponentially large rapid fluctuations of some type to
yield an enormous amount of cancellation.

It is important to understand how this might come about.  There
are two obvious scenarios:  One is that there are nonzero
couplings to essentially all of the mesons but there are very
strong cancellations. The variable $f$, which summarizes the
coupling information, might have a rapidly oscillating phase,
which would serve to make it very small in the average. Another
possibility is that the couplings are zero to all except an
exponentially small fraction of the energetically allowed final
mesons.  It can then have ``natural'' size couplings to this tiny
fraction of states.  At first sight this second scenario might
seem far fetched. However, it is worth recalling {\it how} the
Hagedorn spectrum emerges in an idealized string theory.  It does
not do so with an essentially smooth density of states---with the
states more or less uniformly distributed and a rapidly increasing
density.  Rather, the levels come in {\it highly} degenerate
groups.  The mass of the states in the $n^th$ band of excitations
is given by
\begin{equation}
m_n^2= 2 \pi \sigma n + const
\end{equation}
where $n$ is an integer\cite{Pol}.  Moreover, there is some
evidence for such bands emerging in the real world of $N_c=3$
\cite{Afonin}. The Hagedorn nature of the spectrum is encoded in
the degeneracy of the band which grows exponentially with $n$. Now
suppose hypothetically that when a given hadron in some band, $n$,
decays into two hadrons with fixed quantum numbers in lighter
bands it does so by going into a single hadron of each type.  In
such a scenario the couplings to each state are of natural size
when they are non-zero but are only non-zero for an exponentially
small number of states.  It should be clear, however, that this
scenario, even if fundamentally correct, must set in gradually.
Clearly, for any finite mass state the string theory will not be
ideal and the degenerate bands of states will be split.  It is
{\it a priori} very unlikely that that coupling to the states of
such a split band are restricted to a single state.

In summary, we have shown that the decay widths predicted in a
string description require individual meson coupling constants to
be much larger than the average coupling required by the quark
loop through duality. The arguments we have formulated for both
requirements are based on fairly general properties of the
large-$N_c$ limit.  It remains unclear how relevant these
cancellations are for the $N_c=3$ world. We do not with certainty
know how this cancellation occurs, but one interesting possibility
is that it may be related to the division of the string spectrum
into discrete degenerate bands.

This work is supported by the U. S. Department of Energy under
grant number DE-FG02-93ER-40762.

\section{Appendix}

This appendix gives some of the technical details for the general
arguments presented in the paper.

First we present a specific expression for the amplitude for the
meson process we are comparing the the quark loop with three
insertions, the definition of the spectral strength we use to
compare the meson regime to the quark regime, definitions of the
averages we take, and an exact expression for the meson decay
width. To make the the argument concrete we focus here on some
particular class of meson decays: a process in which a particle of
type $a$ decays into two particles of types $b$ and $c$.  The
quantum number of the various types of mesons are left general in
this; we will have to make appropriate contractions over the
various quantum numbers of the states.  These mesons have the
quantum number associated with some type of local quark bilinear
current. The currents for these particles are represented by
$A^{(\alpha)}$, $B^{(\beta)}$, and $C^{(\gamma)}$. Indices in
parentheses indicate that these particles are tensors of arbitrary
rank. To connect the decay amplitudes  to the currents we consider
the three point function for the currents  $A$, $B$ and $C$:
\begin{widetext}
\begin{equation} \label{eqn:threeptfirst}
\begin{split}
i M & \left(s_0,s_1,s_2,\lambda_0,\lambda_1,\lambda_2\right) =
\\&\sum_{N,M,K} d_M d_N d_K \langle0|A^{(\alpha)}|a(\lambda_0,N)\rangle
          \frac{D_{a(\alpha)(\mu)}}{s_0-m(N)^2}
           \Lambda^{(\mu)(\nu)(\xi)}
                  \frac{D_{b(\beta)(\nu)}}{s_1-m(M)^2}
                   \frac{D_{c(\gamma)(\xi)}}{s_2-m(K)^2}
           \langle b(\lambda_1,M)|B^{(\beta)}|0\rangle
           \langle c(\lambda_2,K)|C^{(\gamma)}|0\rangle
\end{split}
\end{equation}
\end{widetext}
Here, $D_{(\alpha)(\mu)}$ is the tensor form of the propagator
(for some general type of propagator), while we have written
explicitly the pole structure. $D$ will depend on the mass of the
meson, and its form will be determined by the type of meson.
$\Lambda^{(\mu)(\nu)(\xi)}$ is the 3-point vertex whose behavior
we will analyze. $\lambda_0$, $\lambda_1$, and $\lambda_2$
represent the particular state of the particle, including the spin
state. $d_M$, $d_N$, and $d_K$ are the degeneracies of the states.

We make the following definitions:
\begin{equation}
\begin{split}
\langle0|A^{(\alpha)}|a(\lambda_0,N)\rangle & \equiv a_{\lambda_0 N}
     e^{i\phi_a}\epsilon^{(\alpha)}
\\\langle0|B^{(\beta)}|b(\lambda_1,M)\rangle & \equiv b_{\lambda_1 M}
     e^{i\phi_b}\epsilon^{(\beta)}
\\\langle0|C^{(\gamma)}|c(\lambda_2,K)\rangle & \equiv c_{\lambda_2 K}
     e^{i\phi_c}\epsilon^{(\gamma)}
\end{split}
\end{equation}
\begin{equation} \label{eqn:fdef}
\begin{split}
f \equiv & \epsilon^{(\alpha)}\epsilon^{(\beta)}\epsilon^{(\gamma)}
     D_{a(\alpha)(\mu)}D_{b(\beta)(\nu)}D_{c(\gamma)(\xi)}
\\ & \times \Lambda^{(\mu)(\nu)(\xi)} e^{i(\phi_a+\phi_b+\phi_c)}
\end{split}
\end{equation}
Here, $a_{\lambda_0}(m_0)$, $b_{\lambda_1}(m_1)$, and
$c_{\lambda_2}(m_2)$ are defined to be real and positive, and all of
the phase absorbed is into the
$f(m_0,m_1,m_2,\lambda_0,\lambda_1,\lambda_2)$ function.

We want to extract the spectral strength on the meson side, defined as follows:
\begin{widetext}
\begin{equation}
\begin{split}
i \Delta  & \left(s_0,\lambda_0,\dots\right) =
\\ & (2 \pi i)^3 \left(
\sum_{N} a_{\lambda_0 N} d_N \delta(s_0-m(N)^2) \right)
          \left(
\sum_{M} b_{\lambda_1 M} d_M \delta(s_1-m(M)^2) \right)
  \left(
\sum_{K} c_{\lambda_2 K} d_K \delta(s_2-m(K)^2) \right)
        f(s_0,\lambda_0,..)
\end{split}
\end{equation}
\end{widetext}
To do this, we define pole prescriptions for the propogators by making
the replacement $s \to s \pm i \epsilon$.

We then see that the spectral strength is an asymmetric combination of
amplitudes made with these replacements:
\begin{equation}
\begin{split}
i \Delta&\left(s_0,s_1,s_2,\lambda_0\dots\right) =
\\&   i \left[   M\left(s_0+i\epsilon,s_1+i\epsilon,s_2+i\epsilon,\lambda_0\dots\right)\right.
\\&             -M\left(s_0+i\epsilon,s_1+i\epsilon,s_2-i\epsilon,\lambda_0\dots\right)
\left.        + \cdots \right]
\end{split}
\end{equation}
The quark loop amplitude will also be a function of these same $s$'s,
with the precise dependance determined by the choice of frame, so we
should be able to make the same replacements and therefore extract an
expression corresponding to $\Delta\left(s_0,s_1,s_2\right)$. We will
define $Q_3\left(s_0,s_1,s_2\right)$ as the amplitude of a quark loop
with three insertions. $Q_2\left(s\right)$ is similarly defined as the
amplitude of a quark loop with two insertions. $i \Delta_Q$ will then
depend on $Q_3$ in the same way as $i \Delta$ depends on $M$.

We now return to the $\Delta$ function defined above, and switch from
the individual meson values to the average values for the functions
$f$, $a_{\lambda N}$, $b_{\lambda M}$, and $c_{\lambda K}$, defined
as the values they would each take if all four were assumed to be
constant over the integration range.

With $\rho_{a \lambda_0}(s_0)$, $\rho_{b \lambda_1}(s_1)$, and
$\rho_{c \lambda_2}(s_2)$ as the densities of hadron states, we define
these average functions as follows, with $R_0 \equiv \left(s_0,s_0+\Delta
s_0\right)$,$R_1 \equiv \left(s_1,s_1+\Delta s_1\right)$, and
$R_2 \equiv \left(s_2,s_2+\Delta s_2\right)$ defining the small momentum region
over which we are comparing the quark regime to the hadronic regime:
\begin{equation} \label{eqn:rhodef}
\rho(s) \Delta s = \sum_{N \in R}{d_N}
\end{equation}
\begin{equation} \label{eqn:abardef}
   \overline{a_{\lambda_0}}(s_0)  \equiv  \left( \sum_{N, s(N) \in R_0}  a_{\lambda_0 N} \frac{d_N}{\rho_{a
             \lambda_0}(s_0) \Delta s_0}
             \right), etc...
\end{equation}
\begin{widetext}
\begin{equation} \label{eqn:fbardef}
\overline{f}\left(s_0,\lambda_0,\dots\right)  \equiv
             \sum_{
\begin{subarray}{1} N, s(N) \in R_0
                               \\M, s(M) \in R_1
                       \\K, s(K) \in R_2
               \end{subarray}
         }
              f\left(s(N),\lambda_0,\dots\right)
\frac{d_N}{\rho_{a \lambda_0}(s_0) \Delta s_0}
\frac{d_M}{\rho_{b \lambda_1}(s_1) \Delta s_1}
\frac{d_K}{\rho_{c \lambda_2}(s_2) \Delta s_2}
\frac{a_{\lambda_0 N}}{\overline{a_{\lambda_0}}(s_0)}
             \frac{b_{\lambda_1 M}}{\overline{b_{\lambda_1}}(s_1)}
             \frac{c_{\lambda_2 K}}{\overline{c_{\lambda_2}}(s_2)}
\end{equation}
\end{widetext}
Note here that $\frac{d_N}{\rho(s) \Delta s} \to \frac{d_N}{\sum_{N'
\in R}{d_{N'}}}$, which, for a smooth $d_N$, is just equal to one over
the number of states in our small integration range.

The integral over a range of states allows us to equate the two
descriptions as follows:
\begin{equation} \label{eqn:Q3def}
\begin{split}
Q_3' \left(2 \pi i\right)^3 & \equiv i\Delta_Q \left(s_0,\lambda_0,\dots\right) =
\\ & \left(2 \pi i\right)^3 \overline{f}\left(s_0,\lambda_0,..\right)
\left(\rho_{a \lambda_0}(s_0) \overline{a_{\lambda_0}}(s_0)\right)
\\ & \times      \left(\rho_{b \lambda_1}(s_1) \overline{b_{\lambda_1}}(s_1)\right)
   \left(\rho_{c \lambda_2}(s_2) \overline{c_{\lambda_2}}(s_2)\right)
\end{split}
\end{equation}

We can trivially repeat the above calculation for the two-point
function to find the $\overline{a_\lambda}$, $\overline{b_\lambda}$, and
$\overline{c_\lambda}$ functions:
\begin{equation}
\begin{split}
 \overline{a_{\lambda}}(s)  =  \sqrt{\frac{Q_{2 a} (s,\lambda)}{\rho_{a \lambda}(s)}}, etc...
\end{split}
\end{equation}

The amplitude of the decay for an individual meson is:
\begin{equation}
i M(m_0,\lambda_0 \to \{m_f,\lambda_f\}) =
          \epsilon^{(\alpha)} \epsilon^{(\beta)} \epsilon^{(\gamma)}
      \Lambda_{(\alpha)(\beta)(\gamma)}
\end{equation}
with $\Lambda^{(\alpha)(\beta)(\gamma)}$ related to the $f$
function as in (\ref{eqn:threeptfirst}).

We encode the difference between $f$ and its average by defining
$\overline{f} \equiv R f$, and using the fact that $\rho(s) =
\frac{\rho(m)}{2 m}$, the decay width for a particle of mass $m_0$ to
split into particles of masses $m_1$ and $m_2$ in the CM frame is
just:
\begin{widetext}
\begin{eqnarray}
\Gamma(m_0,\lambda_0 \to m_1,\lambda_1;m_2,\lambda_2) & = & |R|^2
             \frac{Q
}{ \rho_{a \lambda_0}(m_0)
                                \rho_{b \lambda_1}(m_1)
                                \rho_{c \lambda_2}(m_2)}
\\\label{eqn:Qdef} Q(s_0,s_1,s_2,\lambda_0,\lambda_1,\lambda_2)& \equiv &
                             \frac{2 m_0 m_1 m_2 p}{2 \pi m_0^2} \left( \frac{\vert Q_3'(s_0,s_1,s_2)\vert^2}
                                {Q_{2a}(s_0) Q_{2b}(s_1) Q_{2c}(s_2)}
                               \right)
              \left| \frac{\epsilon^{(\mu)}
                                      \epsilon^{(\nu)} \epsilon^{(\xi)}}
                        {\epsilon^{(\alpha)} \epsilon^{(\beta)} \epsilon^{(\gamma)}
                D_{a (\alpha)}^{(\mu)}
                D_{b (\beta)}^{(\nu)}
                D_{c (\gamma)}^{(\xi)}}
                         \right|^2
\end{eqnarray}
\end{widetext}
where $p$ is the magnitude of the momentum of either particle relative
to the CM.

\end{document}